# Effect of Functionalized CNT on Nematic anchoring

Jitendra Kumar,[1] Raj K Gupta,[1] Sandeep Kumar[2] and V. Manjuladevi[1,*]

*1 Dept. of Physics, Birla Institute of Technology and Science, Pilani-333031, India*
*2 Raman Research Institute, Sadashivanagar, Bangalore-560080, India*

*E-mail: manjula@bits-pilani.ac.in

Abstract: Nematic phase is the most fundamental mesophase exhibited by most of the rod shaped anisotropic liquid crystalline molecules. Nematics are orientationally ordered fluids whose average orientation direction can be manipulated on application of electric and magnetic fields. Carbon nanotube (CNT), a highly shape anisotropic object can find numerous industrial application because of its interesting electronic and mechanical properties. The self-organizing properties of nematics can be used to align CNTs dispersed in them. We have dispersed functionalized CNTs in nematic liquid crystal and carried out many experimental studies. We will present results of electro-optic switching and dielectric measurements on some CNT-LC dispersion. We have observed that addition of functionalized CNTs in a liquid crystal (LC) has led to improvement in the nematic ordering which is evidential from enhancement in dielectric anisotropy ($\Delta\varepsilon$) measurement. These results indicate that the anchoring energy at alignment layers has been influenced by presence of FCNT in the LC host. The anchoring enhancement can be attributed to $\pi$-$\pi$ electron stacking between the FCNT, LC and the alignment layer.

**INTRODUCTION**

In the Nematic phase, the *calamitic* or rod-shaped organic molecules have no positional order, but the preferred long axes of LC molecules orient on an average along a particular direction called the director. Nematic liquid crystals (NLCs) exhibit optical, dielectric, and magnetic anisotropy like crystals. [1, 2]. On application of electric and magnetic fields the collective orientation of nematic director can be manipulated [1, 2]. Display applications mainly rely on the electric Freedericksz transition, a collective reorientation of the director along the direction of an applied electric field. Carbon nanotube (CNT), another highly anisotropic material has generated enormous interest in the scientific community because of its extraordinary electrical, mechanical and magnetic properties [3, 4]. CNTs because of their high geometric aspect ratio, can act as a better dopant to improve upon the properties of the host LC matrix [5, 6]. Dierking et al have studied the effect of CNT on the alignment and reorientation of the LC under both electric and magnetic fields [7, 8]. The reports on electro-optic response of LCs doped with multiwalled and singlewalled carbon nanotubes indicate an improvement in response time and threshold voltage of the LC-CNT composite compared to that of pure LCs [9, 10]. Functionalised CNTs are better candidates in many proposed applications ranging from nanoelectronics [11] to biomedical drug delivery [12] because of their relatively easier solubility [13]. Diamine functionalized CNT in a polyamide matrix shows a better dispersion and a stronger interfacial adhesion as compared to unfunctionalized CNT [14]. In this paper we report the effect of functionalized CNT on some properties of LC host. We have observed a significant enhancement in both dielectric constant and conductance of NLC-FCNT dispersions compared to that of pure LC.

**EXPERIMENTAL**

We have used E7 (Merck), the most commonly used nematic liquid crystal mixture in the display industry, as the host LC material. Acid functionalized CNT (AFCNT) were purchased from Carbon Solutions Inc. We have prepared dispersions of ~ 0.05 wt% of both of these FCNTs with E7. The dispersions were prepared by dissolving the components in chloroform solvent and sonicating for about an hour. The dispersions were used in the further studies after evaporating the solvent. Indium tin oxide coated glass (ITO) plates were coated with polymide PI2555 (PI) and cured at 250°C. The planar aligned cells were prepared using these rubbed PI coated ITO plates. The LC filled cell was placed between crossed polarizers with the rubbing direction making an angle of 45° with either of the polarizers. Prior to any electro-optic measurements, the doped LC cells were observed under a polarizing microscope. The FCNT doped cells reveal a uniform texture like a pure LC cell indicating a uniform nematic director field between crossed polarizers. Electro-optic switching and dielectric measurements have been carried out on pure LC as well as AFCNT doped LC. The electro-optic measurements were carried out by recording the transmitted intensity from the cell using a photodiode as a function of applied voltage. The capacitance and conductance of the pure and doped LC cells were measured using HIOKI 3522- 50 LCR Hi Tester. Because of the anisotropic nature of the LC molecule the nematic phase exhibits dielectric anisotropy, where components parallel and perpendicular to the molecular long axis are $\varepsilon_{\parallel}$ and $\varepsilon_{\perp}$, respectively. As the maximum voltage accessible for dielectric measurements was 5 V, we treat the dielectric permittivity measured at 5 V to be $\varepsilon_{\parallel}$ and that measured at 0.2 V which is less than the Freedericksz transition threshold voltage (Vth) to be $\varepsilon_{\perp}$. The dielectric anisotropy ($\Delta\varepsilon$) is estimated using the difference between the measured values of $\varepsilon_{\parallel}$ and $\varepsilon_{\perp}$. All the measurements were carried out at room temperature (~24°C). The thickness of the planar aligned cells is ~4.5 μm.

**RESULTS & DISCUSSION**

The transmittance as a function of applied voltage at 2.1 kHz is shown in Figure 1. Interestingly, the Vth is found to be ~0.7 V for pure LC as well as AFCNT doped LC. Using the transmitted intensity data, the value of birefringence ($\Delta\mu$) at 24°C is estimated [15]. The value of $\Delta\mu$ is found to be ~ 0.22 for pure LC as well as FCNT doped LC which is in good agreement with previous measurements on pure LC [16, 17]. The dielectric constant ($\varepsilon$) as a function of applied voltage is shown in Figure 2. There is an enhancement in the of FCNT doped LC compared to that of pure LC. The values of $\Delta\varepsilon$ are ~ 9.4, ~ 10.1, respectively for pure LC, and AFCNT doped LC. Russell et al [18] have shown that CNTs induce alignment on the NLC director field along their long axes due to LC-CNT anchoring effect. So, the enhanced $\Delta\varepsilon$ observed in doped LC can be attributed to the LC-FCNT anchoring at the alignment layer. The conductivity ($\sigma$) of AFCNT doped LC is enhanced considerably compared to that of pure LC at all applied voltages as shown in Figure 3. The conductivity of the LC and FCNT doped LC as a function of frequency at a voltage of 0.2 V is shown in Figure 4. The value of conductivity in the low frequency regime (<10 Hz) of FCNT doped LC is less compared to that of pure LC. This indicates that the effect of ions in AFCNT doped LC cells is less compared to that of pure LC. For frequencies above 100 Hz, the conductivity of AFCNT doped LC is enhanced by a factor of 3 compared to that of pure LC. The imaginary part of the $\varepsilon$ ($\varepsilon^{\parallel}$) is calculated using the conductivity data [15]. The frequency variation of imaginary part of $\varepsilon_{\perp}$ ($\varepsilon_{\perp}^{\parallel}$) at an applied voltage of 0.2 V is shown in Figure 5. The relaxation frequencies of pure LC, and AFCNT doped LC are found to be 3 Hz, and 40 Hz respectively. The dielectric spectra of liquids in the frequency range $10^{-2}$ to 10 Hz are determined by the near electrode processes such as the charge exchange between the electrode and charge carriers present in the LC [19]. Whereas the dielectric spectra in the middle frequency region (10- $10^5$ Hz) characterizes bulk properties of the sample. Interestingly, the relaxation frequency of AFCNT doped LC were found to be 40Hz. This indicates that the increase in conductivity of FCNT doped LC in the region above 10 Hz is not due to effect of ions, but due to the presence of FCNT at the alignment layer. Using a dielectric spectroscopy measurement carried out on single walled and multiwalled CNTs, Basu et al. [20] have reported that CNTs have a relaxation frequency in the range of 16 to 80Hz. They report that it is related to space charge polarization, which may involve a mechanism of charge built up at electrode-sample interface due to the presence of CNTs at the electrode. Hence, the observed relaxation frequency of 40 Hz for AFCNT doped LC can be attributed to the presence of AFCNT at the alignment layer. A strong interaction associated with the AFCNT alignment mechanism is mainly due to surface anchoring. The increase in anchoring energy is believed to be from the $\pi - \pi$ electron stacking between CNT, surface alignment layer and LC molecules [21, 10]. The FCNT presents in LC matrix follow the orientation of the surrounding LC molecules which enhances the orientational order in the medium. The dielectric anisotropy is a measure of order parameter of the medium. Thus, the enhancement in $\Delta\varepsilon$ of FCNT doped LC indicates an enhancement in the nematic order. The expression for threshold voltage is, $V_{th} = \pi\sqrt{K_{11}/\varepsilon_0\Delta\varepsilon}$ where $K_{11}$ is the splay elastic constant and $\varepsilon_0$ is dielectric permittivity of free space [1, 10]. As $V_{th}$ of both pure LC and AFCNT doped LC are similar, the elastic constant $K_{11}$ of doped LC is enhanced compared to that of pure LC.

In order to probe the possibility of the presence of AFCNT in alignment layer, we estimated the conductances of ITO coated glass, PI coated ITO, and the films of both pure as well as AFCNT doped LC deposited on this PI coated ITO substrates. The films of pure LC and AFCNT doped LC are deposited by spin coating for 30 minutes at 3000 rpm. We have measured the tunneling current ($I_t$) and topographic heights simultaneously in the spreading resistance imaging (SRI) [22] mode of atomic force microscope (AFM,

NTMDT, Solver Pro). The bias voltage between the diamond coated conducting tips and the substrate was maintained at 0.2 V.

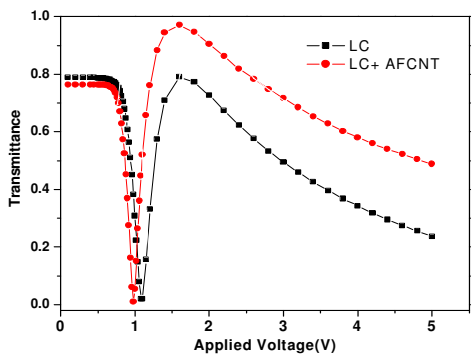
Fig. 1

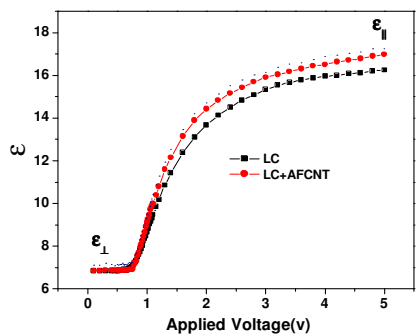
Fig. 2

Figure1: Transmittance of pure LC and FCNT doped LC as a function of applied voltage at 2.1 kHz.
Figure2: Voltage dependent variation of dielectric constant ($\varepsilon$) of pure LC and FCNT doped LC at 2.1 kHz.

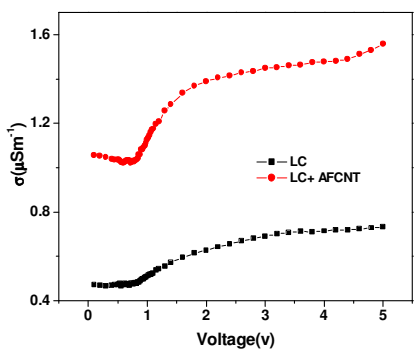
Fig. 3

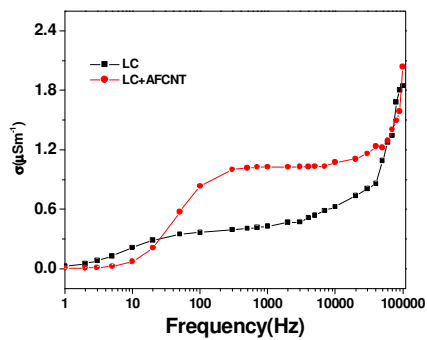
Fig. 4

Figure 3: Voltage dependent variation of conductivity ($\sigma$) of pure LC and FCNT doped LC at 2.1 kHz.
Figure 4: Conductivity ($\sigma$) of pure LC and FCNT doped LC as a function of frequency.

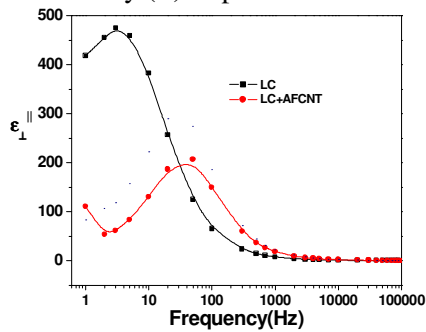
Fig. 5

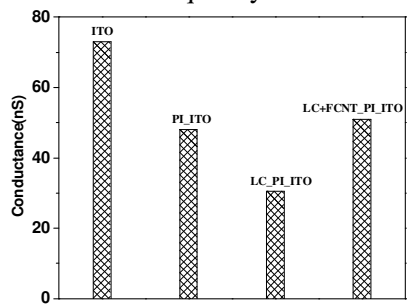
Fig. 6

Figure 5: Imaginary part of dielectric constant conductivity ($\varepsilon_\perp^{||}$) of pure LC and FCNT doped LC as a function of frequency.
Figure 6: Average Conductivity ($\sigma$) of pure LC and AFCNT doped LC estimated using AFM.

The average value of conductance estimated using the tunneling current image profile is shown in figure 8. The value of conductance (G) decreases with the deposition of PI on the ITO substrate. The value of G decreases further with the deposition of LC on such PI treated ITO. Interestingly, the value of G increases on depositing AFCNT doped LC as compared to that of pure LC on PI treated ITO plates. The increase in conductance in AFCNT doped LC is due to incorporation of ODFCNT in the alignment layer in the vicinity of ITO. This confirms the presence of AFCNT in the PI alignment layer.

## CONCLUSION

In conclusion, we have observed that addition of acid functionalized CNTs in a LC has led to improvement in the nematic ordering which is evidential from enhancement in $\Delta\varepsilon$. The observed relaxation frequency of 40 Hz in AFCNT doped LC cell suggests the presence of AFCNT at the alignment layer. The AFM study confirms the AFCNT present at the alignment layer which can lead to better nematic anchoring which further enhances the nematic ordering.

**Acknowledgements:**
The author, V. Manjuladevi gratefully acknowledges DST, India for the support provided under SERC Fast track scheme (SR/FTP/PS-10/2009). Authors also gratefully acknowledge UGC India for DRS-SAP.